\documentclass[aps, pra, 10pt, amsmath, amssymb, superscriptaddress, twocolumn]{revtex4-2}
\usepackage{physics,esint,graphicx,bbm,amsthm}
\usepackage[dvipsnames,svgnames,table]{xcolor}
\usepackage[unicode=true,
  pdfusetitle,
  bookmarks=true,
  bookmarksnumbered=false,
  bookmarksopen=false,
  breaklinks=true,
  pdfborder={0 0 0},
  backref=false,
  colorlinks=true,
hypertexnames=false]{hyperref}
\hypersetup{linkcolor=NavyBlue,urlcolor=NavyBlue,citecolor=NavyBlue}

\newcommand{\mcF}{\mathcal{F}}

\newcommand*{\TT}{{\mkern-1.5mu\mathsf{T}}}

\begin{document}

\title{Revisit on quantum parameter estimation approach for Mach-Zehnder interferometry}
% \title{Phase Difference Estimation Scheme in a Mach-Zehnder Interferometer}

\author{Bing-Shu Hu}
\affiliation{School of Sciences, Hangzhou Dianzi University, Hangzhou 310018, China}
\author{Xiao-Ming Lu}
\email{lxm@hdu.edu.cn}
\affiliation{School of Sciences, Hangzhou Dianzi University, Hangzhou 310018, China}
\affiliation{Zhejiang Key Laboratory of Quantum State Control and Optical Field Manipulation, Hangzhou Dianzi University, Hangzhou 310018, China}

%% Abstract
\begin{abstract}
  The Mach-Zehnder interferometer is a fundamental tool for measuring phase shifts between two light paths, serving as a crucial prototype for achieving high-precision measurements in various scientific and technological applications.
  In this study, we analyze different models for estimating relative phase shift in a general two-arm Mach-Zehnder interferometer.
  We demonstrated that single-parameter estimation models can be reduced from the two-parameter estimation model by imposing appropriate constraints on the parameter space.
  To make quantum Fisher information of the single-parameter estimation models meaningful, the corresponding constraints must be guaranteed in the experiment implementation.
  Furthermore, we apply the quantum Fisher information approach to analyze the Mach-Zehnder interferometer with the an input state composed of a displaced squeezed vacuum state and a coherent state, providing insights into the precision limits of such configurations.
\end{abstract}

\maketitle

\section{Introduction}
Quantum parameter estimation theory is a fundamental tool for understanding the quantum limit of precision in estimating unknown parameters~\cite{Helstrom1967,Helstrom1968,Holevo2011,Braunstein1994,Braunstein1996,Paris2009}.
In cases where the quantum system depends on multiple parameters, the quantum Fisher information (QFI) matrix can be used to analyze the quantum limit of precision in parameter estimation~\cite{Liu2019}.
However, quantum multiparameter estimation is more complicated than single-parameter estimation, not only because of the incompatibility problem~\cite{Ragy2016,Lu2021} but also because of the diversity of parametrization.
Even when focusing on a single parameter of interest, the presence of additional parameters may influence the overall precision of the  estimation process.
To reveal the practical quantum limits on parameter estimation, quantum multiparameter estimation theory should be properly used.

% Classical parameter estimation theory is originated at the beginning of the 20th century, Fisher analyzed the statistical problems in parameter estimation and discussed the criteria for estimators~\cite{Fisher1925,Fisher1922}.
% Rao introduced how it can be used to estimate unknown parameters to obtain more accurate estimation results and the lower bound~\cite{Rao1945}.
% Quantum parameter theory was proposed in the latter half of the 20th century by Helstrom~\cite{Helstrom1967, Helstrom1968, Helstrom1969} and Holevo~\cite{Holevo1973,Holevo2011}.eria for estimators~\cite{Fisher1925,Fisher1922}.
% In 1968, Helstrom discussed the quantum-mechanical form of the Cram\'er-Rao inequality~\cite{Helstrom1968}, and in 1974, Helstrom and Kennedy explored the possibility of simultaneous measurements of noncommutative observables in a quantum receiver~\cite{Helstrom1974}.
% Quantum parameter estimation can attain a precision impossible to a purely classical system~\cite{Caves1981,Bondurant1984}.Helstrom1969} and Holevo~\cite{Holevo1973,Holevo2011}.
% In 1968, Helstrom discussed the quantum-mechanical form of the Cram\'er-Rao inequality~\cite{Helstrom1968}, and in 1974, Helstrom and Kennedy explored the possibility of simultaneous measurements of noncommutative observables in a quantum receiver~\cite{Helstrom1974}.
% Quantum parameter estimation can attain a precision impossible to a purely classical system~\cite{Caves1981,Bondurant1984}.

The Mach-Zehnder interferometer (MZI) is a fundamental optical device that plays a crucial role in studying the coherence properties of light fields~\cite{Ludwig1891}.
A key parameter in this interferometer is the relative phase difference between its two arms, which has been the focus of significant research efforts~\cite{Demkowicz-Dobrzanski2015}.
The QFI approach has been widely utilized to optimize the MZI for various input states and to explore the quantum limit of phase estimation~\cite{Liu2017,Yu2018,Liu2013}.
Notably, Jarzyna and Demkowicz-Dobrza\'{n}ski in Ref.~\cite{Jarzyna2012} pointed out potential flaws in the simplified models of MZI when using the QFI to reveal the fundamental limit of phase estimation, suggesting that these models might not fully reflect the practical limits of phase estimation.

% In 2005, Hradil \textit{et al.} calculated the QFI for inputting two Fock states in Mach-Zehnder interferometer~\cite{Hradil2005}.
% In 2012, Jarzyna \textit{et al.} discussed the QFI for the input state consisting of squeezed vacuum and coherent states in a Mach-Zehnder interferometer~\cite{Jarzyna2012}.
% In this paper,  we will discuss three different schemes for estimating phase and calculates the corresponding QFI.
% In 2024, Abdellaoui \textit{et al.} introduced the QFI for three different schemes of phase estimation in a Mach-Zehnder interferometer with the input state being an entangled coherent state with a vacuum state~\cite{Abdellaoui2024}.

In this paper, we revisit the quantum parameter estimation approach for the MZI.
We will analyze different models for estimating the relative phase shift in the MZI and show how to understand the applicabilities of these models by quantum multiparameter estimation theory.
We will show that the QFI matrix, which is the core quantity in the quantum multiparameter estimation theory, contains enough information about the precision of phase estimation in the MZI.
Even when we are only interested in the relative phase shift between the two arms, the QFI matrix should be used to analyze the precision of the phase estimation.
The QFI matrix can be applied in two ways: (i) to analyze the estimation precision of the relative phase shift in the presence of nuisance parameters and (ii) to analyze the precision of phase estimation in the presence of prior constraints on the parameters.
Several single-parameter models for estimating the relative phase shift in the MZI should be interpreted as the QFI about the relative phase shift in the presence of proper constraints; otherwise, the QFI may be overoptimistic.
Moreover, we will specifically examine the QFI approach for the MZI with an input state composed of a displaced squeezed vacuum state and a coherent state.

% In many practical experiments, the MZI is used to sense and measure tiny phase shifts caused by a signal in various situation, e.g.~gravitational wave detection~\cite{LIGO2009}.
% For such cases, the MZI is operated at a certain working point (i.e., optimal phase point) to achieve the best sensitivity.
% Some phase information can be assumed known when the signal is absent.

This paper is organized as follows.
In Sec.~\ref{sec:general_analysis}, we give a general analysis of phase estimation in a MZI.
In Sec.~\ref{sec:displaced_squeezed_state}, we study the phase estimation in a MZI with a displaced squeezed vacuum state and a coherent state as input.
We summarize our work in Sec.~\ref{sec:conclusion} and give the details of some calculations in the Appendix.

\section{Phase estimation in the MZI}
\label{sec:general_analysis}

\begin{figure}[tb]
  \centering
  \includegraphics{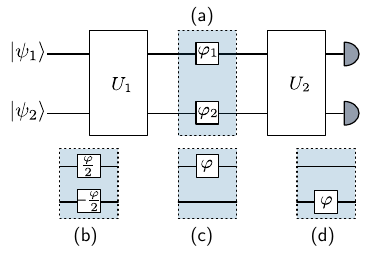}
  \caption{ Different parametrization models of the MZI.
    (a) Two-parameter estimation model. (b, c, d) Various single-parameter estimation models.
  }
  \label{fig:MZI_normal}
\end{figure}

Let us denote by \(a_1\) and \(a_2\) the annihilation operators of the two input modes of an MZI.
The input state is a product state \(\ket{\psi_0}=\ket{\psi_1}\otimes\ket{\psi_2}\).
It first evolves through a unitary process \(U_1\) and then passes through a channel with two arms.
The upper and lower arms cause phase shifts \(\varphi_1\) and \(\varphi_2\), respectively, as shown in Fig.~\ref{fig:MZI_normal}(a).
The corresponding unitary transformation is given by
\begin{equation}
  U_{\varphi_1\varphi_2} = \exp(-i\varphi_1 a_1^\dagger a_1 - i\varphi_2 a_2^\dagger a_2).
\end{equation}
For an optical interferometer, the quantum state typically goes through another unitary process \(U_2\), e.g., a beam splitter,  before the final measurement.
The unitary operator \(U_1\) and \(U_2\) are independent of the phase parameters \(\varphi_1\) and \(\varphi_2\).
The output state \(\ket{\psi}\) is given by \(\ket{\psi}=U_2 U_{\varphi_1\varphi_2} U_1 \ket{\psi_0}\).

The relative phase shift \(\varphi_1-\varphi_2\) between the two arms is usually the parameter of interest.
Therefore, it will be convenient to use \(\varphi_\pm := \varphi_1 \pm \varphi_2\) as the independent parameters instead of the original ones \(\varphi_1\) and \(\varphi_2\).
As the photon number operator \(N := a^\dagger_1 a_1 + a^\dagger_2 a_2\) commute with \(a_1^\dagger a_1 - a_2^\dagger a_2\), the unitary operator \(U_{\varphi_1\varphi_2}\) can be decomposed into
\begin{align}
  U_{\varphi_1\varphi_2}
  =\exp(-\frac{i\varphi_+}{2}N) \exp[-\frac{i\varphi_-}{2} (a_1^\dagger a_1 - a_2^\dagger a_2)].
\end{align}
According to the quantum parameter estimation theory, we can calculate the QFI matrix \(\mcF\) about \(\varphi_\pm\) to analyze the precision of the phase estimation.
For pure states, the QFI matrix \(\mcF\) can be calculated as
\begin{align}\label{eq:QFI_pure_state}
  \mathcal{F}_{jk}
  =4\Re(\braket{\partial_{j}\psi}{\partial_{k}\psi}
  -\bra{\partial_{j}\psi}\ket{\psi}\bra{\psi}\ket{{\partial_{k}\psi}})
\end{align}
with \(\partial_1 = \partial/\partial{\varphi_+}\) and \(\partial_2 = \partial/\partial{\varphi_-}\).
The inverse of the QFI matrix \(\mcF^{-1}\) bounds the covariance matrix \(\mathcal E\) of any unbiased estimator~\cite{Helstrom1967,Helstrom1968} as
\begin{align} \label{eq:qcrb_ineq}
  \mathcal E \geq \nu^{-1}\mcF^{-1},
\end{align}
where \(\nu\) is the number of repetitions of the experiment.

When focusing on the relative phase shift $\varphi_-$, it is tempting to use quantum single-parameter estimation theory for precision analysis.
The analysis often employs various simplified models for estimating the relative phase shift $\varphi_-$ in a MZI, as demonstrated in
Figs.~\ref{fig:MZI_normal} (b), (c), and (d).
However, such approaches must be grounded in robust theoretical frameworks and carefully examined to avoid being overoptimistic about the estimation precision.
In what follows, we shall investigate several existing single-parameter estimation models for the MZI and show how to understand their applicabilities by quantum multiparameter estimation theory.

The typical theoretical models of estimating the relative phase shift are illustrated in Fig.~\ref{fig:MZI_normal}.
With a slight abuse of notation, we use \(\varphi\) instead of \(\varphi_-\) to simplify the notation, as we focus on the relative phase shift.
For the model shown in Fig.~\ref{fig:MZI_normal}(a), the value of the common phase shift \(\varphi_+\) is unknown and not of our interest, that is to say, \(\varphi_+\) is a nuisance parameter~\cite{VanTrees2013}.
The model shown in Fig.~\ref{fig:MZI_normal} (b) assumes that the phase shifts in the two arms are equal in magnitude but opposite in sign, i.e., \(\varphi_1 = {\varphi}/{2}\) and \(\varphi_2 = -{\varphi}/{2}\).
In this model, any common phase shift \(\varphi_+\) is considered irrelevant and is set to zero.
This approach is widely used in optical interferometry~\cite{Yurke1986,Demkowicz-Dobrzanski2015}, especially when the observable finally measured commutes with the total photon number operator \(N\).
The models shown in Fig.~\ref{fig:MZI_normal} (c) and (d) assume that the phase shift occurs only in one arm of the interferometer, while the other arm has no phase shift.

The QFIs about the relative phase shift for these four models are given by
\begin{align}
  \mcF_{\varphi}^{(a)} &= \mcF_{22} - {\mcF_{12}^2} / {\mcF_{11}}, \label{eq:QFI_a}\\
  \mathcal{F}^{(b)}_{\varphi} &= \mathcal{F}_{22}, \label{eq:QFI_b}\\
  \mathcal{F}^{(c)}_{\varphi} &= \mathcal{F}_{11} + 2\mathcal{F}_{12} + \mathcal{F}_{22}, \label{eq:QFI_c}\\
  \mathcal{F}^{(d)}_{\varphi} &= \mathcal{F}_{11} - 2\mathcal{F}_{12} + \mathcal{F}_{22}. \label{eq:QFI_d}
\end{align}
All the above results are expressed in terms of the QFI matrix elements \(\mcF_{jk}\) defined in Eq.~\eqref{eq:QFI_pure_state}, meaning that the two-parameter estimation model shown in Fig.~\ref{fig:MZI_normal}(a) is the most general scenario for estimating the relative phase shift in an MZI and the two-parameter QFI matrix contains sufficient information about the estimation precision.

Note that  \(\mathcal{F}^{(a)}_{\varphi}\) is equal to \(1/(\mcF^{-1})_{22}\) due to the explicit form of the inverse matrix of any \(2 \times 2\) matrix.
It can be seen from Eq.~\eqref{eq:qcrb_ineq} that the variance of any unbiased estimator \(\hat\varphi\) must obey \(\operatorname{Var}(\hat{\varphi}) \geq (\mathcal F^{-1})_{22} / \nu = 1/(\nu \mathcal{F}^{(a)}_{\varphi}) \).
In other words, \(\mathcal{F}^{(a)}_{\varphi}\) is the QFI about the relative phase shift in the presence of nuisance parameter \(\varphi_+\).

\begin{figure}[tb]
  \includegraphics{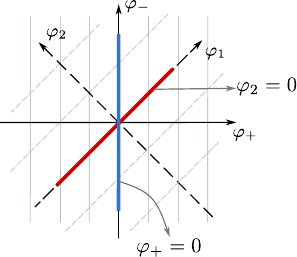}
  \caption{Various constraints on the parameter space.}
  \label{fig:MZI_parameter_constraint}
\end{figure}

The quantities \(\mathcal{F}^{(b)}_{\varphi}\), \(\mathcal{F}^{(c)}_{\varphi}\), and \(\mathcal{F}^{(d)}_{\varphi}\) are the QFIs for single-parameter model of estimating the relative phase shift.
These QFIs were studied in the prior work, e.g., see Ref.~\cite{Jarzyna2012,Ataman2020}.
We here show that they can be obtained from the two-parameter QFI matrix \(\mcF\) by setting proper constraints on the parameter space \((\varphi_1,\varphi_2)\).
Without loss of generality, assume there is a constraint \(f(\varphi_-,\varphi_+)=0\) on the unknown parameters.
The constraint reduces the dimension of the parameter space.
As shown in Fig.~\ref{fig:MZI_parameter_constraint}, the constraint \(f(\varphi_-,\varphi_+)=0\) can be viewed as a curve in the parameter space.
To reduce the model from two-parameter estimation to single-parameter estimation, we can choose \((f,g)\) as the two new parameters, where \(g\) is another independent function of \(\varphi_+\) and \(\varphi_-\).
The QFI matrix in the new basis \(\tilde{\theta}=(f,g)\) is given by
\begin{align} \label{eq:QFI_transform}
  \tilde{\mathcal{F}}
  &= \mathcal J^\TT \mathcal{F}\mathcal J,
\end{align}
where
\begin{equation}
  \mathcal J =
  \mqty(
    \pdv*{\varphi_+}{f}, \pdv*{\varphi_+}{g} \\
    \pdv*{\varphi_-}{f}, \pdv*{\varphi_-}{g}
  )
\end{equation}
is the Jacobian matrix of the parameter transform.
The QFI about the parameter \(g\) with the constraint \(f=c\), where \(c\) is any constant, is given by \(\tilde{\mcF}_{22}\) at \(f=c\).

From the above approach to reduce the model from two-parameter estimation to single-parameter estimation, we can see that \(\mathcal F^{(b)}_\varphi\) is the QFI about the relative phase shift in the presence of the constraint \(\varphi_+=c\).
To derive \(\mathcal F^{(c)}_\varphi\) and \(\mathcal F^{(d)}_\varphi\) from the two-parameter QFI matrix \(\mathcal F\), we need to transform the QFI matrix from the basis \((\varphi_+,\varphi_-)\) to the parameter basis \((\varphi_1,\varphi_2)\) according to Eq.~\eqref{eq:QFI_transform}, that is,
\begin{align}
  \tilde{\mathcal{F}}
  &= \mqty(\mathcal{F}_{11}+2\mathcal{F}_{12}+\mathcal{F}_{22}
    & \mathcal{F}_{11}-\mathcal{F}_{22}\\
    \mathcal{F}_{11}-\mathcal{F}_{22}
  & \mathcal{F}_{11}-2\mathcal{F}_{12}+\mathcal{F}_{22}).
\end{align}
Therefore, \(\mathcal F_\varphi^{(c)}\) is given by \(\tilde{\mathcal F}_{11}\) at \(\varphi_2=c\) and \(\mathcal F_\varphi^{(d)}\) is given by \(\tilde{\mathcal F}_{22}\) at \(\varphi_1=c\).

% To study this scheme, we transform the QFI matrix from the basis \(\theta=(\varphi_+,\varphi_-)\) to the basis \(\tilde{\theta}=(\varphi_1,\varphi_2)\) as follows:
% \begin{align}
%   \tilde{\mathcal{F}}
%   &= \mathcal J^\TT \mathcal{F}\mathcal J,\nonumber\\
%   &= \mqty(\mathcal{F}_{11}+2\mathcal{F}_{12}+\mathcal{F}_{22}
%     & \mathcal{F}_{11}-\mathcal{F}_{22}\\
%     \mathcal{F}_{11}-\mathcal{F}_{22}
%   & \mathcal{F}_{11}-2\mathcal{F}_{12}+\mathcal{F}_{22}),
% \end{align}
% where \(\mathcal J\) is the Jacobian matrix with the entries \(\mathcal J_{ij}:={\partial\theta_i}/{\partial\tilde{\theta}_j}\).
% Hence, the QFI about \(\varphi\) with constraint \(\varphi_2=0\) is
% \begin{align}
%   \mcF^{(c)}_{\varphi}
%   = \mathcal{F}_{11}+2\mathcal{F}_{12}+\mathcal{F}_{22}.
% \end{align}

The above analysis suggests that, in order to make the single-parameter QFIs about the relative phase shift meaningful, the corresponding constraints must be ensured in the experiment.
Otherwise, the single-parameter QFIs might be overoptimistic for the quantum limit of estimation precision.
In many realistic scenarios, e.g., gravitational wave detection~\cite{LIGO2009}, the MZI is operated at an adjustable working phase point and the parameter to be estimated is a tiny phase perturbation.
The model in Fig.~\ref{fig:MZI_normal} (b), (c), (d) stand for different sensing mechanisms in the MZI.
They can be understood as different curves determined by corresponding constraint in the parameter space, as shown in Fig.~\ref{fig:MZI_parameter_constraint}.

Note that \(\mcF^{(c)}_{\varphi}\) is greater than \(\mcF_{22}\) when \(\mcF_{12}\geq 0\).
The reason is not only that the constraint brings in the prior information about the values of parameters but also that there is a scaling operation in the parameter transform from \((\varphi_+,\varphi_-)\) to \((\varphi_1, \varphi_2)\).
In other words, the determinant of the Jacobian matrix \(\mathcal J\) is not equal to 1 so that the determinant of the QFI matrix \(\tilde{\mcF}\) is not equal to the determinant of the original QFI matrix \(\mcF\).
This fact should be taken into account when comparing the QFIs in different models.

From the above analysis, it can be seen that the unconstrained scheme as Fig.~\ref{fig:MZI_normal}(a) shown is suitable when the common phase shift is treated as a nuisance parameter.
Other single-parameter models are not equivalent and might underestimate the Cram\'er-Rao bound on the estimation error of the relative phase shift in the absence of proper constraints.
When there is a constraint on the parameter space, these single-parameter models can be used to analyze the achievable limit of the estimation precision.
We will use these QFIs to analyze the MZI with the displaced squeezed vacuum state and the coherent state as input in next section.

\section{Displaced squeezed vacuum state plus coherent state as input}
\label{sec:displaced_squeezed_state}

\begin{figure}[tb]
  \centering
  \includegraphics{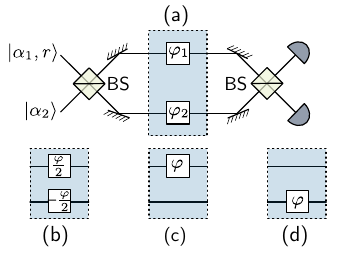}
  \caption{Mach-Zehnder interferometer with a displaced squeezed state and a coherent state as input.}
  \label{fig:MZI_CoSq-Co}
\end{figure}

We now consider a MZI that takes the product state of a displaced squeezed state and a coherent state as the input.
The classical Fisher information for the photon counting of the two output modes under this model is given in Ref.~\cite{Rai2024}.
Here we calculate the QFIs to analyze the fundamental limit of precision in estimating the phase difference between the two arms.
The input state is expressed as
\begin{equation} \label{eq:input_state}
  \ket{\psi_0}=\ket{\alpha_1,r}\otimes\ket{\alpha_2},
\end{equation}
where $\ket{\alpha_1,r}$ is a displaced squeezed vacuum state with a zero squeeze angle, i.e., \(\ket{\alpha_1,r} = D(\alpha_1)S(r)\ket{0}\) with \(D(\alpha_1) = e^{\alpha_1 a^\dagger-\alpha_1^*a}\) and \(S(r) = e^{r(a^2-a^{\dagger2})/2}\) being the displacement and squeeze operators, respectively, and $\ket{\alpha_2}$ is a coherent state.
We henceforth assume that the squeeze parameter \(r\geq 0\).
The two beam splitters are both 50:50 and the corresponding unitary transformations are given by \(U_{\text{BS}} = e^{-i\pi(a^\dagger_1 a_2+a_2^\dagger a_1)/4}\) and \(U_\mathrm{BS}^\dagger\).
After the first beam splitter, the state evolves through a channel with phase shifts \(\varphi_1\) and \(\varphi_2\) in the upper and lower arms, respectively, as shown in Fig.~\ref{fig:MZI_CoSq-Co}(a).
After the second beam splitter, the output state is given by \(\ket{\psi} = U_{\text{BS}}^\dagger U_{\varphi_1\varphi_2} U_{\text{BS}} \ket{\psi_0}\), for which we can calculate the QFI matrix to analyze the precision of the phase estimation.

The MZI has the SU(2) mathematical structure and can be conveniently described by the SU(2) generators~\cite{Yurke1986},
\begin{align}\label{eq:J}
  J_1&=\frac12    \qty(a_1^\dagger a_2+a_2^\dagger a_1),\\
  J_2&=\frac1{2i} \qty(a_1^\dagger a_2-a_2^\dagger a_1),\\
  J_3&=\frac12    \qty(a_1^\dagger a_1-a_2^\dagger a_2),
\end{align}
as well as the total number operator \(N=a_1^\dagger a_1+a_2^\dagger a_2\).
Note that $J_i$ satisfies the SU(2) commutation relation \([J_i,J_j]=i\epsilon_{ijk}J_k\) and $N$ commutes with all $J_i$.
With these operators, the total unitary transformation can be expressed as
\begin{align}
  U_{\text{BS}}^\dagger U_{\varphi_1\varphi_2} U_{\text{BS}} = \exp(-\frac{i}{2} \varphi_+N) \exp(- i \varphi_- J_2),
\end{align}
where \(\varphi_\pm = \varphi_1 \pm \varphi_2\).
Using Eq.~\eqref{eq:QFI_pure_state}, we obtain the elements of the QFI matrix about the parameters \(\varphi_+\) and \(\varphi_-\) as
\begin{align}
  \mathcal{F}_{11} &= \ev*{N^2} - \ev*{N}^2, \nonumber\\
  \mathcal{F}_{12} &= 2\ev*{NJ_2} - 2\ev*{N} \ev*{J_2}, \nonumber\\
  \mathcal{F}_{22} &= 4\ev*{J_2}^2 - 4\ev{J_2}^2, \label{eq:model_F}
\end{align}
and \(\mathcal F_{21}=\mathcal F_{12}\), where the expectation values are taken with respect to the input state \(\ket{\psi_0}\).
Substituting the concrete state \(\ket{\psi_0}=\ket{\alpha_1,r}\otimes\ket{\alpha_2}\), the final result is
\begin{align}
  \mathcal{F}_{11}
  &= |\alpha_2|^2+\frac{\sinh^2 2r}{2} + \Theta_1,\nonumber\\
  \mathcal{F}_{22}
  &=|\alpha_1|^2 +\sinh^2 r + \Theta_2, \nonumber\\
  \mathcal{F}_{12}
  &=-\Im\qty(\alpha_1\alpha_2) \sinh 2r
  +2 \Im\qty(\alpha_1^*\alpha_2) \cosh^2 r,
  \label{eq:QFI_final}
\end{align}
where
\begin{align}
  \Theta_1 &= (\Re\alpha_1)^2e^{-2r} + (\Im\alpha_1)^2e^{2r}, \\
  \Theta_2 &= (\Im\alpha_2)^2e^{-2r} + (\Re\alpha_2)^2e^{2r}.
\end{align}
We give the detailed calculation of the QFI matrix in Appendix A.

% We here consider the relation between the QFI matrix and the resource measured by the mean photon numbers.
% Define \(n_{c1}=|a_1|^2\), \(n_2=|a_2|^2\), and \(n_{1s}=\sinh^2 r\).
% Note that \(n_2\) is the mean photon number of the coherent state \(\ket{\alpha_2}\).
% The mean photon number of the displaced squeezed state \(\ket{\alpha_1,r}\) are \(n_{1c}+n_{1s}\).
% We interpret \(n_{1s}\) and \(n_{1c}\) as the photon number caused by the squeezing operation and the displacement operation, respectively.
% For the cases that both \(\alpha_1\) and \(\alpha_2\) are real, the QFI matrix is simplified as
% \begin{align}
%   \mathcal{F}_{11}
%   &=n_{1c} e^{-2r} + n_2 + \sqrt{n_{1s}(1 + n_{1s})}, \nonumber\\
%   \mathcal{F}_{12}
%   &=0,\nonumber\\
%   \mathcal{F}_{22}
%   & = n_{1c} + n_{1s} + n_2 e^{2r}.
% \end{align}

We taken \(n_1=|\alpha_1|^2\), \(n_2=|\alpha_2|^2\), and the squeeze parameter \(r\) as the measures on  the resources used in the MZI.
In what follows, we will study the optimization of different scenarios over the arguments of \(\alpha_1\) and \(\alpha_2\).

If the common phase shift \(\varphi_+\) is treated as a nuisance parameter, the QFI about \(\varphi_-\) is given by \(\mathcal F^{(a)}_\varphi = \mathcal F_{22} - \mathcal F_{12}^2 / \mathcal F_{11}\), which is smaller than \(\mathcal F_{22}\) unless \(\mathcal F_{12}=0\).
For fixed \(n_1\), \(n_2\), and \(r\), \(\mathcal F_\varphi^{(a)}\) attains its maximum
\begin{equation} \label{eq:F_a_max}
  \mathcal F^{(a)}_\varphi = n_1 + \sinh^2 r + n_2 e^{2r},
\end{equation}
when \(\alpha_1\) and \(\alpha_2\) are both real numbers.
If the common phase shift is treated as a constraint, the QFI \(\mathcal F_\varphi^{(b)}\) about \(\varphi_-\) for fixed \(n_1\), \(n_2\), and \(r\) attains its maximum, which is the same as Eq.~\eqref{eq:F_a_max}, when \(\alpha_2\) is real.

% The single-parameter model shown in Fig.~\ref{fig:MZI_CoSq-Co}(b) is equivalent to the two-parameter model shown in Fig.~\ref{fig:MZI_CoSq-Co}(a) with the constraint \(\varphi_+=0\).
% For such case, the QFI about \(\varphi=\varphi_-\) is \(\mathcal{F}_{\varphi}^{(b)} = \mathcal{F}_{22}\).
% Notice that the QFI matrix given by Eq.~\eqref{eq:model_F} is independent of the values of the common phase shift \(\varphi_+\) and the relative phase shift \(\varphi_-\).
% Therefore, the constraint in fact can be \(\varphi_+=c\) with \(c\) being any constant.
% Note that for general \(\alpha_{1}\), \(\alpha_2\), and \(r\), \(\mcF_{12}\neq 0\), so the single-parameter QFI would given an underestimate bound of the variance of the relative phase difference when the common phase shift is not fixed.

% When the constraint comes to \(\varphi_2=c\) as shown in Fig.~\ref{fig:MZI_CoSq-Co} (c),
% we try another way to calculate the QFI matrix about \(\varphi\).
% For a unitary process \(U\) on initial state \(\ket{\psi_0}\), the QFI matrix about \(\varphi\) can be expressed as~\cite{Liu2019}
% \begin{align}
%   \mathcal{F}&=4\ev{\Delta H}_{\psi_0}
% \end{align}
% with $H:=i\pdv{U}{\varphi}U$.
% Therefore,
When the constraint comes to \(\varphi_2=c\) as shown in Fig.~\ref{fig:MZI_CoSq-Co} (c), the QFI about \(\varphi\) is given by
% \begin{align}
%   \mathcal{F}_{\varphi}^{(c)}
%   &=4\ev*{J_2^2}-4\ev*{J_2}^2+\ev*{N^2}+4\ev*{NJ_2}\nonumber\\
%   &\quad-\ev*{N}^2-4\ev*{N}\!\!\ev*{J_2}.
% \end{align}
\(\mathcal{F}_{\varphi}^{(c)}=\mathcal{F}_{11}+2\mathcal{F}_{12}+\mathcal{F}_{22}\) according to Eq.~\eqref{eq:QFI_c}.
% To analyze the characteristic of this model, we calculate the specific expression of \(\mathcal{F}_{\varphi}^{(c)}\).
We found that if \(\Re \alpha_1=-\Im \alpha_2\)  and \(\Im \alpha_1=\Re \alpha_2\) are satisfied, \(\mcF_\varphi^{(c)}\) has a concise result as follow:
\begin{align}
  \mcF_\varphi^{(c)}=(2+\cosh 2r)\sinh^2 r.
\end{align}
Note that the mean photon numbers of squeezed vacuum state is \(n_s=\sinh^2r\).
Therefore, the expression of QFI is \(\mcF_\varphi^{(c)}=(3 + 2 n_s) n_s\).
% Denote \(n_{c1}, n_{c2}\) are the mean number of photons in the coherent states \(\ket{\alpha_1}\) and \(\ket{\alpha_2}\).
% Assuming \(n_{c1,c2,s} = O(n)\), so the Heisenberg-scaling precision is \(O(1/n)\).
Meanwhile, the variance is bounded from below by
\begin{align}
  \operatorname{Var}(\hat\varphi) \geq \frac1{\mcF_\varphi^{(c)}}=\frac1{(3 + 2 n_s) n_s}.
\end{align}
% But the true scheme in experiment is Fig.~\ref{fig:MZI_CoSq-Co}(a).
% The possibility to achieve the Heisenberg limit is not really to trustable.
% In  Fig.~\ref{fig:MZI_CoSq-Co}(a), the QFI \(\mcF_{\varphi}^{(a)}\) depends on \(\alpha_1\) even through there are some constraint about input state.
% (这里计算了\(\mcF_{\varphi}^{(a)}\) 的结果但是结果不太简洁.)
% So there may be a overestimating operation to achieve the Heisenberg limit.
Similarly, setting \(\Re \alpha_1=\Im \alpha_2\), \(\Im \alpha_1=-\Re \alpha_2\), the QFI also will be \(\mcF_\varphi^{(d)}=(2+\cosh 2r)\sinh^2 r\),
and there will be the same result as \(\mcF_\varphi^{(c)}\).

\section{Conclusion}
\label{sec:conclusion}
In this work, we have analyzed different theoretical models for estimating the relative phase shift in the arms of a MZI.
We pointed out that there are two type of models for estimating the relative phase shift in the MZI: the nuisance parameter model and the constrained model.
The nuisance parameter model treats the common phase shift as a nuisance parameter and the relative phase shift as the parameter to be estimated.
The constrained model sets a constraint on the parameter space to reduce the model from two-parameter estimation to single-parameter estimation.
The selection of the model depends on the experimental setup and the prior information about the parameters.
If the constrained model is used but the constraint is not satisfied in the experiment, the QFI about the relative phase shift might be overoptimistic.
Moreover, we have shown that the QFIs in the constrained model can be obtained from the two-parameter QFI matrix by imposing proper constraints on the parameter space, so for theoretical work the two-parameter QFI matrix is sufficient to analyze the precision of the phase estimation.
Our work provides a unified framework to analyze the precision of phase estimation in the MZI and highlights the importance of the constraints in the parameter space.

\begin{acknowledgments}
  This work is supported by the National Natural Science Foundation of China (Grants No. 12275062 and No. 11935012).
\end{acknowledgments}

\appendix

\section{Calculation of the QFI matrix}
We here present the detailed derivation of the QFI matrix, which is given in Eq.~\eqref{eq:QFI_final}, for the input state \(\ket{\psi_0}=\ket{\alpha_1,r}\otimes\ket{\alpha_2}\) in the MZI.
% To obtain the QFI matrix,  as shown in Eq.~\eqref{eq:model_F}, we need to calculate the expectation values of the operators \(N\), \(J_2\), \(N^2\), \(N J_2\), and \(J_2^2\).

The expectation values of the relevant operator in the displaced squeezed vacuum state \(\ket{\alpha_1,r} = D(\alpha_1)S(r)\ket0\) can be transformed to that in the vacuum state as
\begin{equation} \label{eq:expval_relation}
  \ev*{a_1^{\dagger k} a_1^l } = \ev*{\tilde a_1^{\dagger k} \tilde a_1^l }_0,
\end{equation}
where \(\ev{\bullet}_0\) means that the expectation value is taken in the vacuum state and \(\tilde a_1 = S(r)^\dagger D(\alpha_1)^\dagger a_1 D(\alpha_1) S(r)\) is defined.
The displacement operator \(D(\alpha_1)\) and the squeeze operator \(S(r)\) satisfy the transform relations
\begin{align}
  D(\alpha_1)^{\dagger} a_1 D(\alpha_1) &= a_1 + \alpha_1,\\
  S(r)^{\dagger} a_1 S(r) &= a_1 \cosh r - a_1^\dagger \sinh r.
\end{align}
Therefore, we get
\begin{align} \label{eq:ta1}
  \tilde a_1 &= a_1 \cosh r - a_1^\dagger \sinh r + \alpha_1.
\end{align}
Using Eq.~\eqref{eq:expval_relation} and \eqref{eq:ta1}, it is easy to verify the following expressions for the relevant expectation values:
\begin{align}
  \ev*{a_1} &= \alpha_1, \nonumber\\
  \ev*{a_1^\dagger a_1} &= |\alpha_1|^2+\sinh^2 r, \nonumber \\
  \ev*{a_1 a_1} &= \alpha_1^2-\frac12 \sinh 2r, \nonumber\\
  \ev*{a_1^{\dagger 2} a_1 }  &= 2 \alpha_1^* \sinh^2 r - \frac{\alpha_1}{2} \sinh 2r  + \alpha_1^* |\alpha_1|^2, \nonumber \\
  \ev*{(a_1^{\dagger} a_1)^2}
  &= \qty(|\alpha_1|^2+\sinh^2 r)^2 + \frac12 \sinh^2 2r \nonumber\\
  & \quad + \left\vert \alpha_1 \sinh r - \alpha_1^* \cosh r \right\vert^2.
  \label{eq:ev_displaced_squeezed}
\end{align}
On the other hand, the expectation values of the operators in the coherent state \(\ket{\alpha_2}\) are given by
\begin{align}
  \ev*{a_2} &= \alpha_2, \nonumber\\
  \ev*{a_2^\dagger a_2} &= |\alpha_2|^2, \nonumber \\
  \ev*{a_2 a_2} &= \alpha_2^2, \nonumber\\
  \ev*{a_2^{\dagger 2} a_2 }  &=  \alpha_2^* |\alpha_2|^2, \nonumber \\
  \ev*{(a_2^{\dagger} a_2)^2} &= |\alpha_2|^4 + |\alpha_2|^2.
  \label{eq:ev_coherent}
\end{align}

Now, we are ready to obtain the QFI matrix.
Since the input state of the MZI is a product state, according to Eq.~\eqref{eq:model_F}, \(\mathcal F_{11}\) can be expressed as
\begin{equation}\label{eq:F11}
  \mathcal{F}_{11} =\ev*{(a_1^\dagger a_1)^2} + \ev*{(a_2^\dagger a_2)^2} - \ev*{a_1^\dagger a_1}^2 - \ev*{a_2^\dagger a_2}^2.
\end{equation}
Substituting Eq.~\eqref{eq:ev_displaced_squeezed} and \eqref{eq:ev_coherent} into the above expression, we obtain the expression of \(\mathcal F_{11}\) given in Eq.~\eqref{eq:QFI_final}.
% \begin{align}
%   \mathcal F_{11}
%   &= |\alpha_2|^2+\frac{\sinh^2 2r}{2} + (\Re\alpha_1)^2e^{-2r}+(\Im\alpha_1)^2e^{2r}.
% \end{align}

We now calculate the expression of \(\mathcal F_{22}\).
Substituting \( J_2= (a_1^\dagger a_2-a_2^\dagger a_1)/(2i) \) into Eq.~\eqref{eq:model_F}, we have
\begin{align}
  \mathcal{F}_{22}
  &=-\ev*{a_1^\dagger a_1^\dagger} \!\! \ev*{a_2 a_2}
  -\ev*{a_1a_1} \!\! \ev*{a_2^\dagger a_2^\dagger} \nonumber\\
  & \quad +\ev*{a_1^\dagger a_1} \!\! \ev*{1+a_2^\dagger a_2}
  +\ev*{1+a_1^\dagger a_1} \!\! \ev*{a_2^\dagger a_2} \nonumber\\
  &\quad -|\ev*{a_1^\dagger}\!\! \ev*{a_2} - \ev*{a_1} \!\! \ev*{a_2^\dagger}|^2.
  \label{eq:F22}
\end{align}
Substituting Eq.~\eqref{eq:ev_displaced_squeezed} and \eqref{eq:ev_coherent} into the above expression, we obtain the expression of \(\mathcal F_{22}\) given in Eq.~\eqref{eq:QFI_final}.
% \begin{equation}
%   \mathcal{F}_{22}
%   = |\alpha_1|^2 + \sinh^2 r
%   +(\Re \alpha_2)^2e^{2r}
%   +(\Im \alpha_2)^2e^{-2r}.
% \end{equation}

We now calculate the off-diagonal element of the QFI matrix.
According to Eq.~\eqref{eq:model_F} and \(N=a_1^\dagger a_1 \) and \( J_2= (a_1^\dagger a_2-a_2^\dagger a_1)/(2i) \), it follows that
\begin{align}
  \mathcal{F}_{12}
  % =& 2\ev*{NJ_2} -2 \ev*{N} \ev*{J_2} \nonumber\\
  =& -i\ev*{a_1^\dagger a_1 a_1^\dagger}\!\!\ev*{a_2}
  -i\ev*{a_1^\dagger}\!\!\ev*{a_2^\dagger a_2 a_2} \nonumber\\
  &
  +i \ev*{a^\dagger_1a_1+a_2^\dagger a_2}\!\!\ev*{a_1^{\dagger}}\!\!\ev*{a_2}
  + \mathrm{c.c.},
\end{align}
where \(\mathrm{c.c.}\) means the complex conjugate of the previous terms.
Substituting Eq.~\eqref{eq:ev_displaced_squeezed} and \eqref{eq:ev_coherent} into the above expression, we obtain the expression of \(\mathcal F_{12}\) given in Eq.~\eqref{eq:QFI_final}.

\bibliography{ref.bib}

\end{document}